\documentclass[reprint,amsmath,amssymb,aip,pop,superscriptaddress]{revtex4-2}

\usepackage{graphicx}
\usepackage{dcolumn}
\usepackage{bm}
\usepackage{hyperref}

\begin{document}
\raggedbottom

\title{Equation of state and transport coefficients of warm dense aluminum from mixed deterministic--stochastic density functional theory}
\author{Zi~Li}
\affiliation{\mbox{Institute of Applied Physics and Computational Mathematics, Beijing 100088, China}}
\affiliation{National Key Laboratory of Computational Physics, Beijing 100088, China}
\affiliation{\mbox{Tianfu Institute of Innovative Energy Research, Chengdu, Sichuan 610213, China}}

\author{Weijie~Li}
\affiliation{\mbox{Tianfu Institute of Innovative Energy Research, Chengdu, Sichuan 610213, China}}

\author{Cong~Wang}
\email{wang\_cong@iapcm.ac.cn}
\affiliation{\mbox{Institute of Applied Physics and Computational Mathematics, Beijing 100088, China}}
\affiliation{National Key Laboratory of Computational Physics, Beijing 100088, China}
\affiliation{\mbox{Tianfu Institute of Innovative Energy Research, Chengdu, Sichuan 610213, China}}

\author{Ping~Zhang}
\affiliation{\mbox{Institute of Applied Physics and Computational Mathematics, Beijing 100088, China}}
\affiliation{National Key Laboratory of Computational Physics, Beijing 100088, China}
\affiliation{\mbox{Tianfu Institute of Innovative Energy Research, Chengdu, Sichuan 610213, China}}

\author{Xianjue~Peng}
\affiliation{\mbox{Institute of Applied Physics and Computational Mathematics, Beijing 100088, China}}
\affiliation{\mbox{Tianfu Institute of Innovative Energy Research, Chengdu, Sichuan 610213, China}}

\date{\today}

\begin{abstract}
Aluminum is a reference standard in high-energy-density research and serves as a liner material in megampere Z-pinch facilities.
Using mixed deterministic--stochastic finite-temperature density functional theory, we compute the equation of state and transport coefficients of liquid aluminum up to temperatures of $1000$~eV and compare the results with model-based approaches.
We find that for $T<200$~eV the density-functional results differ from the models by more than 10\% at high density, with a maximum deviation of over 30\% (at $T=10$~eV), leading to significant discrepancies in the Hugoniot curve at high compression; the calculated electrical and thermal conductivities deviate from model predictions by 26\% to 63\% at $T\sim100$~eV. 
We further employ the Drude model and the Epperlein--Haines framework to examine the magnetic-field dependence of the conductivities.
These results can provide valuable input data for radiation-hydrodynamics codes.
\end{abstract}

\maketitle

\section{Introduction}
\label{sec:introduction}

Aluminum is a reference standard in high-energy-density (HED) research.
Its shock response has been measured extensively in plate-impact and pulsed-power experiments~\cite{mitchell1981,knudson2015} and described by widely used theoretical EOS models~\cite{kerley1987,lomonosov2007}.
The metal also calibrates opacity models~\cite{colgan2015}, transport closures~\cite{recoules2005,daligault2016}, and multiphase equations of state~\cite{lomonosov2007,sjostrom2016} in shock physics, as well as in laboratory astrophysics~\cite{perry2000,remington2006}.
Reliable density--temperature $(\rho,T)$ tables for aluminum are required in hydrodynamic calculations that treat the metal as a reference material, tamper, or structural liner.

On megampere Z-pinch facilities, an imploding metallic liner or sleeve figures in several inertial-confinement schemes.
Magnetized liner inertial fusion (MagLIF) uses a cylindrical liner that compresses preheated fuel under an axial magnetic field~\cite{gomez2014}.
Owing to its machinability, chemical simplicity, and well-documented Hugoniot, aluminum has been used as a solid cylindrical liner in MagLIF-related experiments on the Z facility~\cite{sinars2010,sinars2011}.
In these experiments, the liner passes through warm-dense $(\rho,T)$ states.
Radiation-hydrodynamics codes need pressure $P$, specific internal energy~$E$, electrical conductivity~$\sigma$, and thermal conductivity~$\kappa$ on a $(\rho,T)$ mesh to follow that evolution.

Quantum molecular dynamics (QMD) based on finite-temperature Kohn--Sham density functional theory (DFT)~\cite{kohnsham1965,mermin1965} has yielded extensive published \textit{ab initio} data for aluminum at the low-$T$ end of the $(\rho,T)$  mesh~\cite{vocadlo2002,bouchet2009,minakov2014,sjostrom2016,desjarlais2002,mazewet2005,recoules2005,vlcek2012,witte2017}.
Within this framework, the computed $P$ and $E$ in the liquid and shocked regimes agree with shock measurements at moderate compression~\cite{sjostrom2016,minakov2014}, and Kubo--Greenwood $\sigma$ and $\kappa$ agree with experiment for liquid aluminum~\cite{recoules2005,desjarlais2002,vlcek2012}.
Published Kohn--Sham QMD data for liquid aluminum, however, terminate near $T\sim10$eV\cite{sjostrom2016,minakov2014}, a temperature range in which the ionic equation of state (EOS) can still be tied to experiment.

The Kohn--Sham orbital formulation of finite-temperature DFT does not extend to the $100$--$1000$~eV electron temperatures addressed here.
With increasing~$T$, thermal occupation spreads over a long tail of partially filled bands, and convergence of $P$ and~$E$ may require resolving occupations down to $10^{-4}$ or smaller~\cite{blanchet2020}.
The number of thermally occupied bands then grows steeply with~$T$, and diagonalization-based QMD is already impractical by $T\sim100$~eV and offers no practical path to keV tables.
Radiation-hydrodynamics codes therefore rely on Thomas--Fermi--Dirac (TFD) average-atom models~\cite{more1988} or ionization-equilibrium (IONEQ) tables~\cite{macfarlane2006} for electronic thermodynamics, and on Spitzer--H\"arm~\cite{spitzer1953} or Lee--More~\cite{leemore1984} fits for transport.
The extent to which these models deviate from \textit{ab initio} aluminum in the warm-dense-matter regime, and how this discrepancy depends on~$\rho$, are still not well understood.

Mixed deterministic--stochastic finite-temperature DFT~\cite{cytter2018,white2020} samples the Mermin density matrix with a small deterministic Kohn--Sham subset plus stochastic orbitals, without summing explicitly over the full thermal band manifold.
Here, we employ this method to obtain the electronic thermal EOS and to build $(\rho,T)$ tables for warm dense aluminum, and to evaluate the transport coefficients from the Kubo--Greenwood formula.
In the following, the method and computational details are described in Sec.\ref{sec:methods}, and the electronic thermodynamic comparisons, the principal Hugoniot, and the zero-field and magnetized transport results are given in Sec.\ref{sec:results}, followed by a brief summary in Sec.~\ref{sec:conclusion}; the construction of the liquid aluminum EOS is described in the Appendix.

\section{Theory and methods}
\label{sec:methods}

\subsection{Mixed deterministic--stochastic finite-temperature DFT}
\label{sec:methods:mdft}

When the electron temperature rises in warm-dense matter, finite-temperature Kohn--Sham DFT requires a summation over a long tail of partially occupied bands; the standard diagonalization-based approach becomes prohibitive by $T\sim100$~eV.
Stochastic DFT (sDFT)~\cite{cytter2018} introduces a set of random orbitals and evaluates density-matrix traces as statistical averages over a finite stochastic sample, avoiding explicit summation over the full thermal band manifold.
Mixed deterministic--stochastic DFT (mDFT)~\cite{white2020} further splits the electronic states into two parts: a small number of low-lying, strongly occupied Kohn--Sham orbitals are obtained by deterministic diagonalization, while the high-lying, weakly occupied tail is sampled with stochastic orbitals orthogonalized to the Kohn--Sham subspace, balancing accuracy and cost.
Compared with pure sDFT, mDFT treats the dominant occupied contribution exactly and confines stochastic noise to the weakly occupied tail, so that comparable accuracy is reached with fewer random orbitals and lower statistical variance in the thermodynamic averages.

In mDFT, the finite-temperature density matrix or Fermi--Dirac operator is written as a function of the Hamiltonian $\hat{H}$,
\begin{equation}
  \hat{f}_H = \frac{1}{1+\exp[(\hat{H}-\mu)/(k_{\mathrm{B}}T)]}.
  \label{eq:fermi_op}
\end{equation}
The electron density $n(\mathbf{r})$, which enters the Hamiltonian as a functional $\hat{H}=\hat{H}[n]$, is expressed as
\begin{equation}
  n(\mathbf{r}) = 2\sum_{a=1}^{N_\chi}\bigl|\langle\tilde{\chi}_a|\hat{f}_H^{1/2}|\mathbf{r}\rangle\bigr|^2
  + 2\sum_{i=1}^{N_\varphi} f(\varepsilon_i)\,|\psi_i(\mathbf{r})|^2,
  \label{eq:mdft_density}
\end{equation}
where $\tilde{\chi}_a$ and $\psi_i$ are stochastic and Kohn--Sham orbitals, respectively.
Self-consistent iteration between Eqs.~\eqref{eq:fermi_op} and~\eqref{eq:mdft_density}, together with the solution of the Kohn--Sham equations in the low-energy region, determines the finite-temperature equilibrium charge density $n(\mathbf{r},T)$. The corresponding pressure, internal energy, and other thermodynamic averages are then evaluated from this density.

In practice, $\hat{f}_H^{1/2}$ is evaluated through a Chebyshev expansion,
\begin{equation}
  \hat{f}_H^{1/2} \equiv g(\hat{h}) = \sum_{n=0}^{\infty} C_n[g]\,T_n(\hat{h}),
  \qquad \hat{h} = \frac{\hat{H}-\bar{E}}{\Delta E},
  \label{eq:chebyshev}
\end{equation}
where $\bar{E}$ and $\Delta E$ set the spectral window of $\hat{H}$, and $T_n$ and $C_n[g]$ are Chebyshev polynomials and the expansion coefficients, respectively.
Application of $T_n(\hat{h})$ to a stochastic orbital is carried out by a three-term recurrence, so that no full diagonalization is required.

\subsection{Kubo--Greenwood transport coefficients}
\label{sec:methods:transport}

The electrical conductivity~$\sigma$ and thermal conductivity~$\kappa$ could be determined by combining the Kubo linear-response formalism with the mDFT framework~\cite{liu2025altransport}.
Since the mDFT method does not directly yield eigenvalues, the conductivities are derived from the current response function, which is expressed in terms of the trace of the current operator,
\begin{equation}
  C_{nm}(t) = -\frac{i}{\hbar}\,\theta(t)\,
  \mathrm{Tr}\!\bigl\{\hat{f}_H\bigl[\hat{\mathbf{J}}_n,\hat{\mathbf{J}}_m(t)\bigr]\bigr\}.
  \label{eq:kubo_corr}
\end{equation}
Here $[\,,\,]$ denotes the commutator, and $\theta(t)$ is the Heaviside step function.
The operators $\hat{\mathbf{J}}_1$ and $\hat{\mathbf{J}}_2$ represent the electrical and heat currents, respectively.

Using the mixed Kohn--Sham and stochastic orbitals, the trace in Eq.~\eqref{eq:kubo_corr} is evaluated as~\cite{liu2025altransport}
\begin{equation}
  \begin{split}
    T_{nm}(t)
    &= \sum_{a,b}
    \langle\varphi_a^{+}(t/2)|\hat{\mathbf{J}}_n|\varphi_b^{-}(t/2)\rangle \\
    &\quad\times
    \langle\varphi_b^{-}(-t/2)|\hat{\mathbf{J}}_m|\varphi_a^{+}(-t/2)\rangle,
  \end{split}
  \label{eq:mdft_transport_trace}
\end{equation}
with time-evolved states
\begin{align}
  |\varphi^{+}(t/2)\rangle &= e^{i\hat{H}t/(2\hbar)}\hat{f}_H^{1/2}|\varphi\rangle, \label{eq:mdft_phi_plus}\\
  |\varphi^{-}(t/2)\rangle &= e^{i\hat{H}t/(2\hbar)}(1-\hat{f}_H)^{1/2}|\varphi\rangle, \label{eq:mdft_phi_minus}
\end{align}
where $|\varphi\rangle$ denotes a Kohn--Sham or stochastic orbital.
For Kohn--Sham orbitals, the time evolution reduces to phase factors, and no additional propagation is required.

\subsection{Computational details}
\label{sec:methods:details}

The electronic thermal EOS and transport properties of aluminum are computed using the mDFT method~\cite{cytter2018,white2020}, as implemented in the ABACUS code~\cite{liu2022}.
The calculations employ a norm-conserving pseudopotential~\cite{hamann2013} with 13 valence electrons and the Perdew--Burke--Ernzerhof exchange--correlation functional~\cite{perdew1996}.
For the electronic thermal contribution, a cubic cell containing a single aluminum atom is used, with the cell size determined by the specified density; the Brillouin zone is sampled with a $4\times4\times4$ Monkhorst--Pack $k$-point mesh.
The energy cutoff is set to 500--1000~Ry, which keeps the relative error of the free energy below $10^{-4}$.
To isolate the electronic thermal components of free energy and pressure, we subtract the corresponding quantities of a near-zero-temperature ($T=300$~K) system at the same density from the computed finite-temperature values.

The transport properties of aluminum are calculated with the same pseudopotential and functional as above.
The supercell contains from 4 aluminum atoms (for the $T=500$~eV system) to 32 aluminum atoms (for $T=20$~eV system); a $2\times2\times2$ Monkhorst--Pack $k$-point mesh is used, and the energy cutoff is set to 230--400~Ry.
First, 4000 steps of quantum molecular dynamics are performed to obtain disordered structures of aluminum in the warm-dense regime.
Three random configurations are then selected for transport calculations, yielding the frequency-dependent electrical and thermal conductivities.
Finally, the electrical/thermal conductivity curves are averaged and fitted to obtain the zero-field transport coefficients.

\section{Results}
\label{sec:results}

\subsection{Electronic thermal contribution}
\label{sec:results:eos}

\begin{figure*}[t]
  \includegraphics[width=\textwidth]{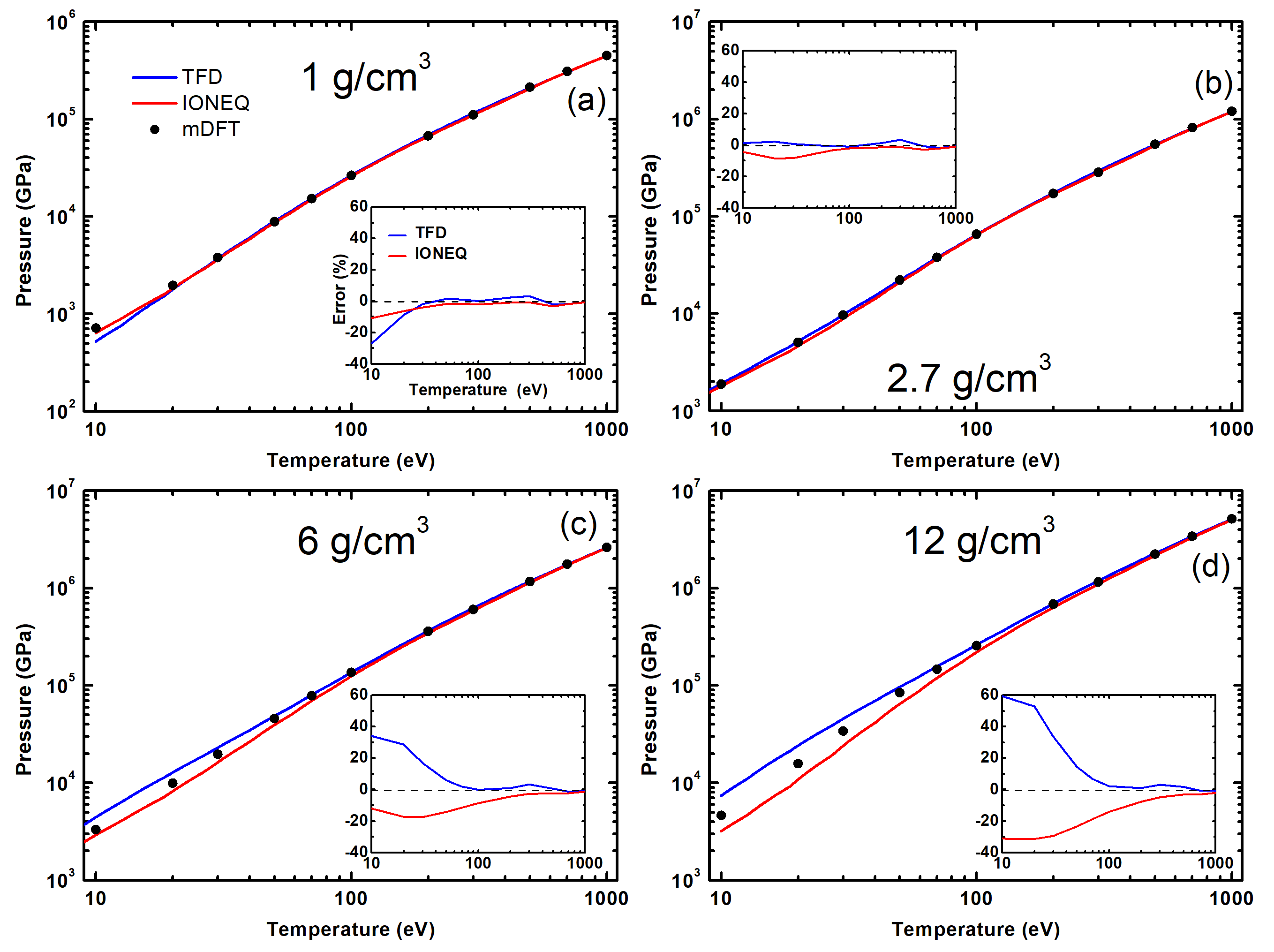}
  \caption{Electronic pressure~$P_{\mathrm{e}}$, from TFD (blue), IONEQ (red) and mDFT (black solid circles)  methods, as a function of temperature~$T$ at different densities: (a) 1 g/cm$^3$, (b) 2.7 g/cm$^3$, (c) 6 g/cm$^3$, and (d) 12 g/cm$^3$.
  Insets show the deviations of the TFD and IONEQ pressures relative to mDFT.}
  \label{fig:P_eos}
\end{figure*}

\begin{figure*}[t]
  \includegraphics[width=\textwidth]{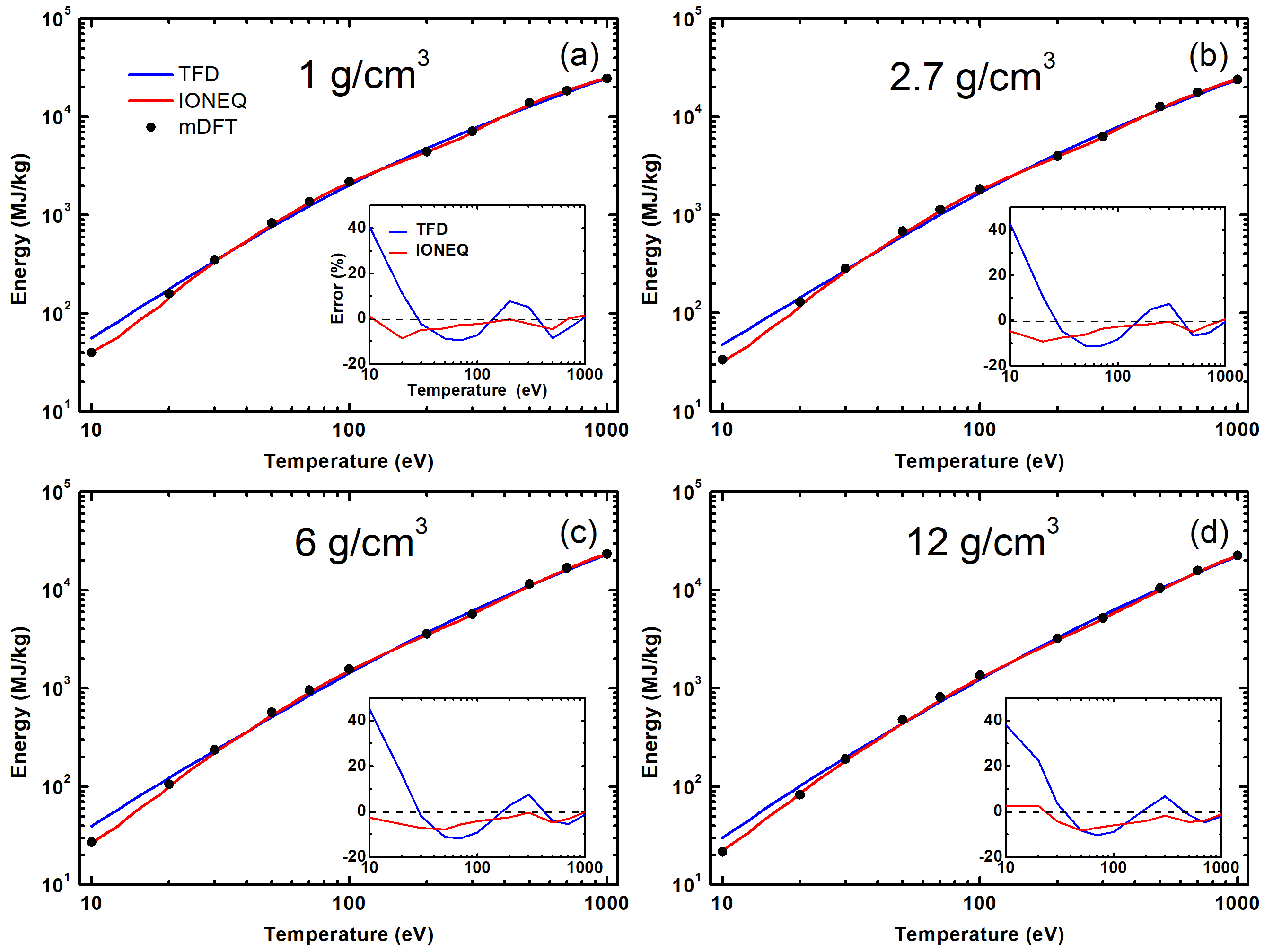}
  \caption{
Electronic specific internal energy $E_{\mathrm{e}}$ from the TFD (blue), IONEQ (red), and mDFT (black solid circles) methods as a function of temperature $T$ at different densities: (a) 1 g/cm$^3$, (b) 2.7 g/cm$^3$, (c) 6 g/cm$^3$, and (d) 12 g/cm$^3$. Insets show the deviations of the TFD and IONEQ internal energies from mDFT.
}
  \label{fig:E_eos}
\end{figure*}

Using mDFT, we compute the electronic thermal pressure~$P_{\mathrm{e}}$ and the specific internal energy~$E_{\mathrm{e}}$ of aluminum for $T=10$--$1000$~eV and $\rho=1$--$18$~g/cm$^3$.
Figs.~\ref{fig:P_eos} and~\ref{fig:E_eos} show the mDFT results for $P_{\mathrm{e}}$ and $E_{\mathrm{e}}$, as functions of temperature at typical densities $\rho=1$, $2.7$, $6$, and $12$~g/cm$^3$. The results from TFD~\cite{more1988} and IONEQ~\cite{macfarlane2006} methods are also shown for comparison. The insets in Figs.~\ref{fig:P_eos} and~\ref{fig:E_eos} show the relative deviations of the TFD and IONEQ results from mDFT, defined as the difference divided by the corresponding mDFT value .

For the electronic thermal pressure~$P_{\mathrm{e}}$, at ambient compression ($\rho=2.7$~g/cm$^3$), both TFD and IONEQ stay within 10\% of mDFT at temperatures of $10$--$1000$~eV.
In the high-density regime ($\rho=6$ and 12~g/cm$^3$), the models also agree within 10\% for $T\ge200$~eV. At lower $T$, TFD and IONEQ results are systematically higher and lower than mDFT, respectively.
Specifically, 
at $\rho=12$~g/cm$^3$ and $T=10$~eV, the pressure deviations are  $+59\%$ for TFD and $-31\%$ for IONEQ.
At low density ($\rho=1$~g/cm$^3$) the deviations are smaller; only at $T=10$~eV does either model exceed 10\%, with both models underestimating the pressure relative to mDFT.
For the internal energy~$E_{\mathrm{e}}$, IONEQ remains within 10\% of mDFT at every density and temperature.
TFD exceeds 10\% for $T<100$~eV, and at $T=10$~eV, the deviation is up to $\sim$45\%.

At extremely high temperatures ($T\sim1000$~eV), the pressures from different methods all approach the ideal-gas values, i.e., $P_{\mathrm{e}}=Nk_{\mathrm{B}}T\rho/m$ with $N$ the number of ionized electrons per aluminum atom ($N=13$ in the extreme high-$T$ limit), $\rho$ the mass density of aluminum, and $m$ the mass of a single aluminum atom. The internal energies are $\sim20\%$ higher than the ideal-gas value $E_{\mathrm{e}}=3Nk_{\mathrm{B}}T/2$, because the ideal-gas model does not include ionization energy. 

\subsection{Principal Hugoniot}
\label{sec:results:hugoniot}

Using the computed electronic thermal contributions to the energy and pressure, we construct a wide-range equation of state for aluminum over an extended liquid domain.
The cold contribution of the EOS is obtained from our low-temperature QMD simulation data, and the ion-thermal contribution is obtained by subtracting the other EOS components from the QMD pressure and energy at representative state points and fitting the residual; details are given in the Appendix.
We also assemble liquid-region EOS tables for aluminum based on the TFD and IONEQ electronic thermal contributions.
In the high-temperature regime examined here ($T>10$~eV), the electronic thermal contribution dominates the total EOS, thus, the curves of pressure and internal energy versus the temperature from the total EOS are similar to those of the electronic contributions in Figs.~\ref{fig:P_eos} and~\ref{fig:E_eos}.

\begin{figure}[t]
  \includegraphics[width=\linewidth]{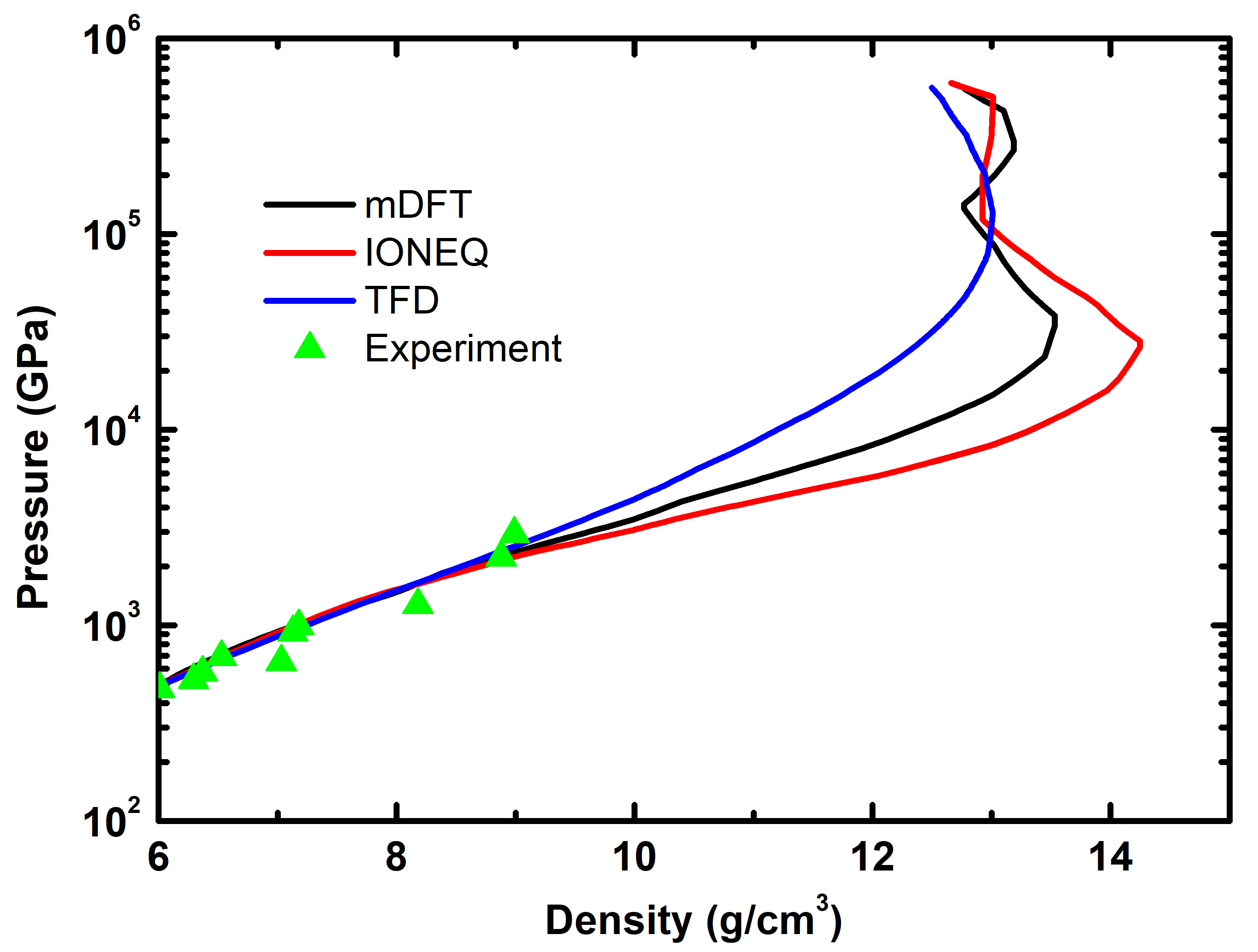}
  \caption{Principal Hugoniot of aluminum based on EOS tables constructed with TFD (blue), IONEQ (red), and mDFT (black) electronic contributions. Experimental data are shown as green solid triangles.}
  \label{fig:hugoniot}
\end{figure}

With the wide-range equation-of-state parameters in hand, the principal Hugoniot of aluminum can be calculated.
The principal Hugoniot is obtained from the Rankine--Hugoniot energy jump~\cite{duvall1977},
\begin{equation}
  E - E_0
  = \tfrac{1}{2}\bigl(P+P_0\bigr)
    \left(\frac{1}{\rho_0}-\frac{1}{\rho}\right),
  \label{eq:hugoniot}
\end{equation}
where 0 indicates the initial state, set as the ambient condition, $\rho_0=2.7$~g/cm$^3$ and $T_0=300$~K; the reference pressure $P_0$ and specific internal energy $E_0$ are obtained from the EOS.
For each trial compression density $\rho$, the temperature $T$ is found by solving Eq.~(\ref{eq:hugoniot}), with $E=E(\rho,T)$ and $P=P(\rho,T)$ from EOS data; both the $T$--$\rho$ and $P$--$\rho$ Hugoniot curves can then be obtained.
Figure~\ref{fig:hugoniot} compares the wide-range Al Hugoniot obtained with our constructed EOS data, where the electronic thermal contribution is obtained from mDFT, TFD, or IONEQ.

For densities $\rho<8$~g/cm$^3$, all three curves coincide and rise monotonically with compression. 
At $\rho\approx8$~g/cm$^3$, TFD and IONEQ give pressures of $1519$~GPa and $1536$~GPa, respectively, both close to the mDFT result of $1488$~GPa (relative difference below 5\%).

As $\rho$ increases further, the pressure gap widens rapidly. At $\rho=10$~g/cm$^3$, TFD (IONEQ) gives a  pressure of $4385$~GPa ($3069$~GPa), about 26\% higher (12\% lower) than the mDFT value of 3470~GPa. At $\rho=12$~g/cm$^3$, the TFD and IONEQ pressures are 120\% higher and 31\% lower than the mDFT value.
These deviations track the 10--60\% differences in electronic thermal pressure noted in Sec.~\ref{sec:results:eos}.

For densities above 12~g/cm$^3$, the Hugoniot curve develops multiple turning points.
The mDFT curve exhibits two local maxima in the density, at $\rho=13.53$ and 13.19~g/cm$^3$, corresponding to compression ratios of 5.01 and 4.88, respectively.
The IONEQ curve shows the compression maximum at higher density, $\rho=14.25$~g/cm$^3$ (ratio 5.28), consistent with its lower pressure relative to mDFT at high compression; a second IONEQ compression maximum at $\rho=12.94$~g/cm$^3$ (ratio 4.79) is not obvious. In contrast, for TFD, only one compression maximum is observed at $\rho=12.94$~g/cm$^3$, owing to the absence of electronic shell structure in the TFD model.

Experimental Hugoniot data for aluminum~\cite{simonenko1985,ragan1982,avrorin1987,ragan1984,trunin1995jetp,trunin1995ht} are also shown in Fig.~\ref{fig:hugoniot}. The available experimental data mainly lie below $9$~g/cm$^3$. Below this density, the experimental and theoretical results are in good agreement. Since the theoretical Hugoniots are also close to one another in this range, the experiments cannot discriminate among them.

\subsection{Electrical and thermal conductivity}
\label{sec:results:transport}

We then examine the transport properties using the mDFT method. 
The frequency-dependent electrical and thermal conductivities are first obtained. The zero-field coefficients $\sigma_0$ and $\kappa_0$ are then obtained by fitting these conductivities with the Drude formula 
\begin{equation}
  \sigma(\omega)=\frac{\sigma_0}{1+\omega^2\tau_\sigma^2},
  \label{eq:drude}
\end{equation}
where $\tau$ is the relaxation time. $\kappa_0$ is fitted from the $\kappa(\omega)$ following the same formula.

\begin{figure}[t]
  \includegraphics[width=\linewidth]{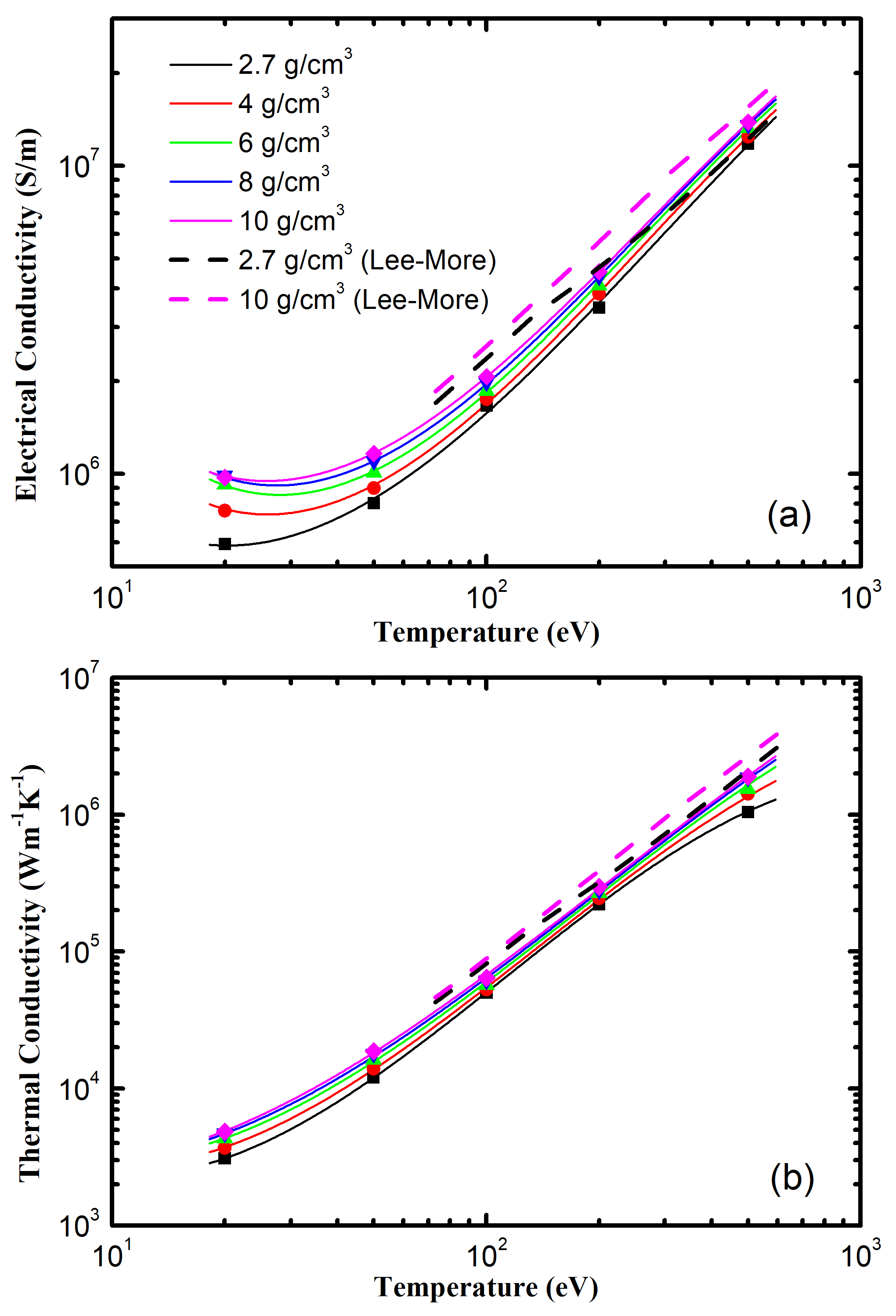}
  \caption{
Zero-field transport coefficients as a function of $T$ at densities $\rho=2.7$--10~g/cm$^3$ (symbols): (a) electrical conductivity $\sigma_0$, and (b) thermal conductivity $\kappa_0$. The solid lines indicate the polynomial fits, and the dashed lines are the results from the Lee--More model at $\rho=2.7$ and $10$~g/cm$^3$.
}
  \label{fig:sigma-kappa}
\end{figure}

The obtained $\sigma_0$ and $\kappa_0$ for densities from 2.7g/cm$^3$ to 10g/cm$^3$ and temperatures from 20eV to 500eV are shown in Fig.\ref{fig:sigma-kappa}, and the electrical relaxation times fitted from $\sigma(\omega)$ are listed in Table\ref{tab:tau_sigma}. 
At each density, $\sigma_0$ rises monotonically with~$T$. Typically, for $\rho=2.7$g/cm$^3$, $\sigma_0$ increases by a factor of 20 from $T=20$eV ($\sigma_0=5.9\times10^5$S/m) to $T=500$eV ($\sigma_0=1.2\times10^7$S/m). In contrast, the electrical conductivity increases only slightly with~$\rho$, by less than a factor of 2 from $\rho=2.7$g/cm$^3$ to $\rho=10$g/cm$^3$. 
The thermal conductivity $\kappa_0$ also rises monotonically with~$T$ but more significantly, by a factor of $\sim$300 from 20eV to 500eV. The electrical relaxation time $\tau$ increases with~$T$ but decreases with~$\rho$. 
The results for the electrical and thermal conductivities at 2.7g/cm$^3$ are consistent with previous mDFT calculations~\cite{liu2025altransport}. 
The transport coefficients from the Lee--More model~\cite{leemore1984} are also shown in Fig.\ref{fig:sigma-kappa}. They show similar dependences on temperature and density, but the values deviate from the mDFT results, especially at lower temperature. For example, the electrical (thermal) conductivity deviates from the mDFT result by 42\% (63\%) and 26\% (38\%) at $T=100$eV for densities of 2.7g/cm$^3$ and 10g/cm$^3$, respectively.

To facilitate their use in Radiation-hydrodynamics codes, the obtained transport coefficients are fitted to a bivariate polynomial in density and temperature. Owing to the extended $\rho$--$T$ domain and the large variation of the transport coefficients, all quantities entering the fit are taken in logarithmic form, with the highest powers of $\lg\rho$ and $\lg T$ set to 2 and 3, respectively: \begin{align}
 \lg\sigma_0 &= \sum_{m=0}^{2}\sum_{n=0}^{3} c_{m,n}(\lg\rho)^m(\lg T)^n,\\ 
 \lg\kappa_0 &= \sum_{m=0}^{2}\sum_{n=0}^{3} d_{m,n}(\lg\rho)^m(\lg T)^n. 
\label{eq:sigma_fit} 
\end{align}
Here, $\lg(x)$ denotes $\log_{10}(x)$, and the units of density, temperature, electrical conductivity, and thermal conductivity in $\lg(x)$ are g/cm$^3$, eV, S/m, and W\,m$^{-1}$\,K$^{-1}$, respectively.
The fitting coefficients $c_{m,n}$ and $d_{m,n}$ are listed in Tables~\ref{tab:sigma_coeff} and~\ref{tab:kappa_coeff}, and the fit curves are shown in Fig.~\ref{fig:sigma-kappa}, with a mean relative error of 1.3\% (2.4\%) and a maximum relative error of 5.4\% (6.8\%) for the electrical (thermal) conductivity.
Note that the transport coefficients from the mDFT calculations are obtained for $\rho=2.7$--10~g/cm$^3$ and $T=20$--500~eV; thus, the fitting formulas are valid only within this regime.

\begin{table}[t]
  \caption{
Electrical relaxation time $\tau_\sigma$ (fs) obtained by fitting the dynamic conductivity with the Drude model, for densities $\rho=2.7$--10g/cm$^3$ and temperatures $T=20$--500eV.
}
  \label{tab:tau_sigma}
  \begin{ruledtabular}
  \begin{tabular}{c|ccccc}
    $T$ (eV) & 2.7~g/cm$^3$ & 4~g/cm$^3$ & 6~g/cm$^3$ & 8~g/cm$^3$ & 10~g/cm$^3$ \\
    \hline
    20 & 0.088 & 0.088 & 0.082 & 0.075 & 0.058 \\
    50 & 0.100 & 0.074 & 0.049 & 0.037 & 0.024 \\
    100 & 0.158 & 0.118 & 0.087 & 0.070 & 0.057 \\
    200 & 0.260 & 0.203 & 0.154 & 0.128 & 0.110 \\
    500 & 0.790 & 0.555 & 0.413 & 0.389 & 0.333 \\
  \end{tabular}
  \end{ruledtabular}
\end{table}

\begin{table}[b]
  \caption{Fitting coefficients $c_{m,n}$ for the electrical conductivity.}
  \label{tab:sigma_coeff}
  \begin{ruledtabular}
  \begin{tabular}{c|cccc}
    $m\backslash n$ & 0 & 1 & 2 & 3 \\
    \hline
    0 & $  1.222$ & $  5.206$ & $-2.046$ & $  0.336$ \\
    1 & $ 22.960$ & $-31.515$ & $ 14.294$ & $-2.125$ \\
    2 & $-14.700$ & $ 20.562$ & $-9.383$ & $  1.398$ \\
  \end{tabular}
  \end{ruledtabular}
\end{table}

\begin{table}[b]
  \caption{Fitting coefficients $d_{m,n}$ for the thermal conductivity.}
  \label{tab:kappa_coeff}
  \begin{ruledtabular}
  \begin{tabular}{c|cccc}
    $m\backslash n$ & 0 & 1 & 2 & 3 \\
    \hline
    0 & $  8.262$ & $-11.015$ & $  7.108$ & $-1.255$ \\
    1 & $-3.888$ & $ 10.323$ & $-7.033$ & $  1.443$ \\
    2 & $  0.295$ & $-2.454$ & $  2.187$ & $-0.512$ \\
  \end{tabular}
  \end{ruledtabular}
\end{table}

\subsection{Magnetic-field effects on $\sigma$ and $\kappa$}
\label{sec:results:magnetic}
Next, we examine the transport properties under a magnetic field, which is useful for numerical simulations of megampere Z-pinch facilities. In a magnetized plasma, transport along the magnetic field is essentially unchanged, whereas the perpendicular components are strongly suppressed. The magnetic response can be described approximately with the classical Drude model, where the perpendicular components of the electrical and thermal conductivities are scaled by a factor of $1/(1+\chi^2)$. Here, Hall parameter $\chi=|e|B\tau/m_e$, with $B$ the magnetic field, $e$ and $m_e$ the electron charge and mass, and $\tau$ the electrical relaxation time (Table~\ref{tab:tau_sigma}) obtained by fitting the frequency-dependent electrical conductivity.

\begin{figure}[t]
  \includegraphics[width=\linewidth]{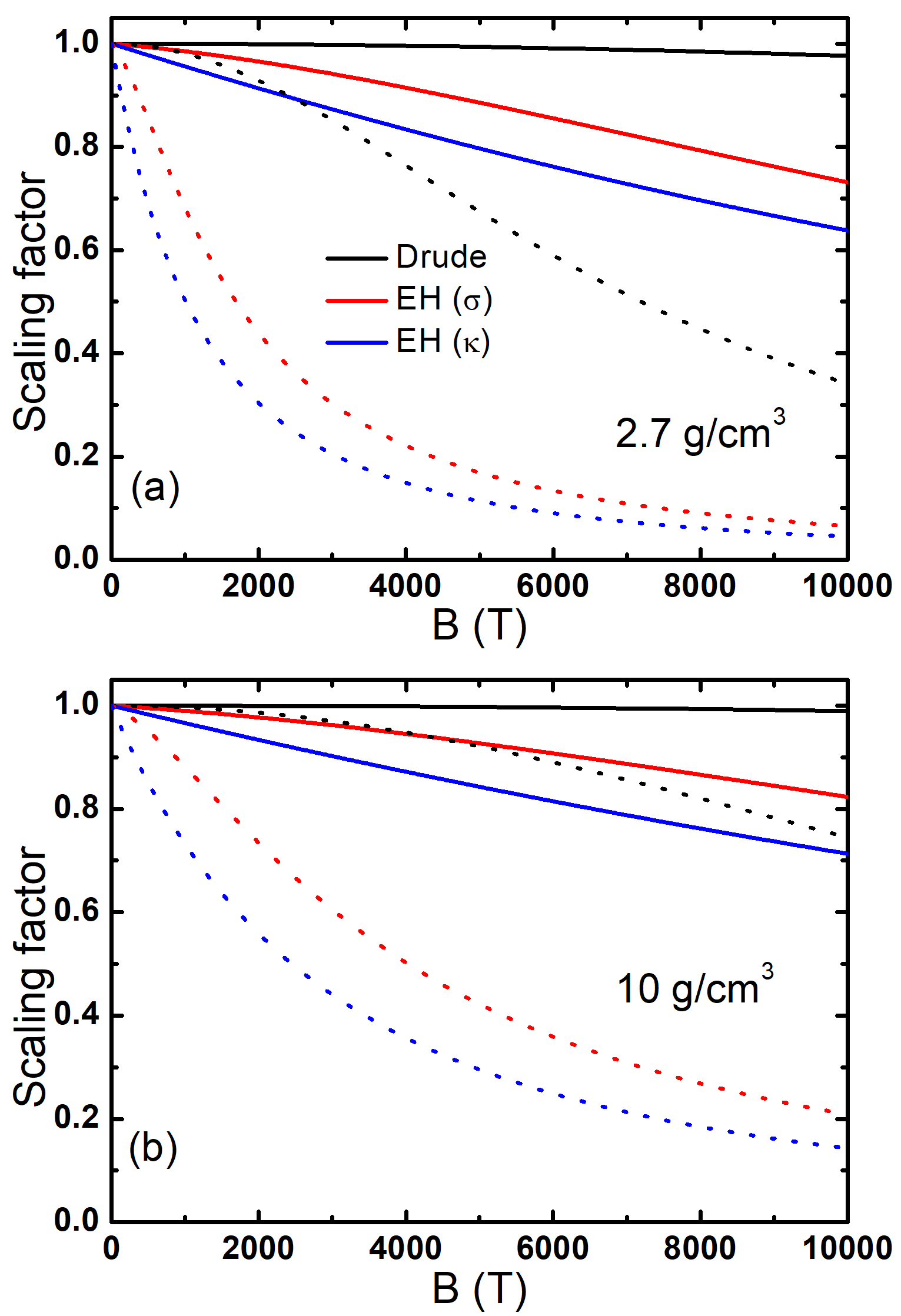}
  \caption{ 
Scaling factors as a function of magnetic field~$B$ from the Drude model (black) and the Epperlein--Haines framework for electrical (red) and thermal (blue) conductivities, at $T=20$~eV (solid) and $500$~eV (dotted). (a) At a density of $2.7$~g/cm$^3$, and (b) at a density of $10$~g/cm$^3$.
}
  \label{fig:magneto}
\end{figure}

Figure~\ref{fig:magneto} shows these Drude scaling factors versus~$B$ at $\rho=2.7$ and $10$~g/cm$^3$ for $T=20$~eV (black solid) and $500$~eV (black dotted). Because $\tau$ is short at 20~eV, even a magnetic field of $10^4$~T leaves the Drude factor close to unity (0.98 for 2.7~g/cm$^3$ and 0.99 for 10~g/cm$^3$); at 500~eV the Drude ratios fall to 0.34 and 0.74.

Epperlein and Haines~\cite{epperlein1986} (EH) computed the magnetized
electron transport coefficients of a fully ionized plasma by direct
numerical solution of the linearized Fokker--Planck equation, and provided
rational-function fits in the Hall parameter~$\chi$, with
coefficients tabulated for different ion charge~$Z$.
For a given~$Z$, the dimensionless Braginskii coefficients~\cite{braginskii1958}
are obtained from these rational-function fits; the electrical
and thermal conductivity anisotropy ratios (or scaling factors)
are then evaluated from the Braginskii coefficients.

In Fig.~\ref{fig:magneto}, the scaling factors from the EH framework
are shown as red (for $\sigma$) and blue (for $\kappa$) curves, with $Z$ taken as the ionization of
aluminum obtained from the TFD model~\cite{more1988}.
The EH scaling factors saturate at large~$B$, unlike the $1/(1+\chi^2)$ decay in the Drude model. Furthermore, for magnetic fields below $10^4$~T, the scaling factors of the EH model are significantly smaller than those of the Drude model. For example, at $\rho=2.7$~g/cm$^3$, $T=500$~eV, and $B=10^4$~T, the electrical and thermal scaling factors from the EH framework are 0.06 and 0.04, respectively, which are only 19\% and 13\% of the corresponding Drude value (0.34).

\section{Summary}
\label{sec:conclusion}

We employed the mDFT method to construct the equation of state
and evaluate the transport coefficients of liquid aluminum at
temperatures up to 1000~eV.
For $T<200$~eV, the electronic thermal contribution to the aluminum EOS from
mDFT differs from the models by more than 10\% at high density, with a maximum deviation of over 30\% (at $T=10$~eV).
Based on this electronic thermal contribution, we constructed
the EOS of aluminum in the liquid region and obtained the principal
Hugoniot.
For densities below 8~g/cm$^3$, the Hugoniot curves from mDFT and the
models essentially coincide and also agree with the experimental data; in the density range of 8--12~g/cm$^3$,
the pressure difference grows and exceeds 10\% at 10~g/cm$^3$.
At still higher densities, these differences in the electronic thermal EOS cause the compression maxima on the TFD and IONEQ Hugoniots to differ
significantly from those of mDFT.

We further calculated the electrical and thermal conductivities of
aluminum at high temperatures using mDFT and found that they deviate from
model predictions by 26\% to 63\% at $T\sim100$~eV.
To facilitate the use of transport parameters in hydrodynamic
simulations, we fitted the electrical and thermal conductivities as
functions of density and temperature.
We also examined the magnetic-field dependence of the electrical and thermal conductivities using the Drude model and the EH framework.
At strong magnetic fields ($\sim 10^4$~T), the transport coefficients perpendicular to the magnetic field are substantially smaller than those parallel to it.

\begin{acknowledgments}
This work was supported by the Science Challenge Project (No.~TZ2025013) and the Funding of National Key Laboratory of Computational Physics (No.~6142A05240101).
\end{acknowledgments}

\appendix*

\section{Construction of the liquid aluminum EOS}
\label{sec:app:eos}

We focus on the liquid region of aluminum and represent the EOS by the standard split of the specific Helmholtz free energy~$F$ into a cold term $F_{\mathrm{c}}$, an ionic thermal term $F_{\mathrm{ion}}$, and an electronic thermal term $F_{\mathrm{e}}$~\cite{sjostrom2016,wu2021},
\begin{equation}
  F(\rho,T) = F_{\mathrm{c}}(\rho) + F_{\mathrm{ion}}(\rho,T) + F_{\mathrm{e}}(\rho,T).
  \label{eq:Fsplit}
\end{equation}
$F_{\mathrm{c}}$ describes the $T\to 0$ limit, $F_{\mathrm{ion}}$ carries the thermal motion of nuclei, and $F_{\mathrm{e}}$ accounts for electronic excitation and ionization.
The electronic thermal contribution enters through $P_{\mathrm{e}}$ and $E_{\mathrm{e}}$, which are already obtained in the main text and then interpolated onto a dense $(\rho,T)$ grid.
The cold term $F_{\mathrm{c}}$ and the ionic thermal term $F_{\mathrm{ion}}$ are determined from a series of Kohn--Sham quantum molecular dynamics simulations covering the liquid region over a broad $(\rho,T)$ range, from which we extract the total pressure $P_{\mathrm{tot}}^{\mathrm{(QMD)}}(\rho,T)$.

The cold contribution is represented by the Vinet--Rose equation~\cite{vinet1989},
\begin{equation}
  F_{\mathrm{c}}(V) = F_0 + \frac{4V_0 B_0}{(B_0'-1)^2}
  \left[1-(1+X)\mathrm{e}^{-X}\right],
  \label{eq:vinet}
\end{equation}
with $X=3(B_0'-1)[(V/V_0)^{1/3}-1]/2$,
where $V_0$ is the specific volume at zero pressure and zero temperature, and $B_0$, $B_0'=(\mathrm{d}B/\mathrm{d}P)_{T}$, and $F_0$ are the isothermal bulk modulus, its pressure derivative, and the cold Helmholtz free energy at $V_0$, respectively.
The cold pressure is $P_{\mathrm{c}}=-(\partial F_{\mathrm{c}}/\partial V)_{T}$.
At each density $\rho_i$ in the fitting window, $P_{\mathrm{tot}}^{\mathrm{(QMD)}}$ along the available finite-$T$ isochores (above the melting point) is linearly extrapolated to $0$~K to obtain $P_{\mathrm{c}}(\rho_i)$.
A least-squares fit to the set $\{P_{\mathrm{c}}(\rho_i)\}$ obtained at different densities $\rho_i$, with specific volume $V=1/\rho$, gives $\rho_{\mathrm{ref}}=2.297$~g/cm$^3$ ($V_0=1/\rho_{\mathrm{ref}}$), $B_0=43.25$~GPa, and $B_0'=4.955$.
$F_0$ is an additive constant and does not affect the construction of this single-phase liquid EOS.

The ionic thermal free energy for the liquid region is written as~\cite{wu2021}
\begin{equation}
  F_{\mathrm{ion}} = F_{\mathrm{ion}}^{\mathrm{Debye}} + F_{\mathrm{cell}},
  \label{eq:Fion_split}
\end{equation}
where the Debye term is
\begin{equation}
  F_{\mathrm{ion}}^{\mathrm{Debye}}(T,V)
  = k_{\mathrm B}T\left[3\ln\bigl(1-\mathrm{e}^{-\theta(V)/T}\bigr)
  - \Phi\!\left(\frac{\theta(V)}{T}\right)\right],
  \label{eq:ion_debye}
\end{equation}
with $\Phi(y)=(3/y^{3})\int_{0}^{y}x^{3}[\mathrm{e}^{x}-1]^{-1}\mathrm{d}x$, and $F_{\mathrm{cell}}$ is the finite-cell correction.
The Debye temperature $\theta(V)$ is obtained from the piecewise liquid Gr\"uneisen parameter~\cite{sjostrom2016} $\gamma(\rho)$ through $\mathrm{d}\ln\theta=\gamma\,\mathrm{d}\ln\rho$.
The ionic thermal pressure $P_{\mathrm{ion}}=-(\partial F_{\mathrm{ion}}/\partial V)_{T}$ is least-squares fitted to the QMD pressures after subtracting the cold and electronic contributions.

%


\begin{thebibliography}{45}%
\makeatletter
\providecommand \@ifxundefined [1]{%
 \@ifx{#1\undefined}
}%
\providecommand \@ifnum [1]{%
 \ifnum #1\expandafter \@firstoftwo
 \else \expandafter \@secondoftwo
 \fi
}%
\providecommand \@ifx [1]{%
 \ifx #1\expandafter \@firstoftwo
 \else \expandafter \@secondoftwo
 \fi
}%
\providecommand \natexlab [1]{#1}%
\providecommand \enquote  [1]{``#1''}%
\providecommand \bibnamefont  [1]{#1}%
\providecommand \bibfnamefont [1]{#1}%
\providecommand \citenamefont [1]{#1}%
\providecommand \href@noop [0]{\@secondoftwo}%
\providecommand \href [0]{\begingroup \@sanitize@url \@href}%
\providecommand \@href[1]{\@@startlink{#1}\@@href}%
\providecommand \@@href[1]{\endgroup#1\@@endlink}%
\providecommand \@sanitize@url [0]{\catcode `\\12\catcode `\$12\catcode
  `\&12\catcode `\#12\catcode `\^12\catcode `\_12\catcode `\%12\relax}%
\providecommand \@@startlink[1]{}%
\providecommand \@@endlink[0]{}%
\providecommand \url  [0]{\begingroup\@sanitize@url \@url }%
\providecommand \@url [1]{\endgroup\@href {#1}{\urlprefix }}%
\providecommand \urlprefix  [0]{URL }%
\providecommand \Eprint [0]{\href }%
\providecommand \doibase [0]{https://doi.org/}%
\providecommand \selectlanguage [0]{\@gobble}%
\providecommand \bibinfo  [0]{\@secondoftwo}%
\providecommand \bibfield  [0]{\@secondoftwo}%
\providecommand \translation [1]{[#1]}%
\providecommand \BibitemOpen [0]{}%
\providecommand \bibitemStop [0]{}%
\providecommand \bibitemNoStop [0]{.\EOS\space}%
\providecommand \EOS [0]{\spacefactor3000\relax}%
\providecommand \BibitemShut  [1]{\csname bibitem#1\endcsname}%
\let\auto@bib@innerbib\@empty
\bibitem [{\citenamefont {Mitchell}\ and\ \citenamefont
  {Nellis}(1981)}]{mitchell1981}%
  \BibitemOpen
  \bibfield  {author} {\bibinfo {author} {\bibfnamefont {A.~C.}\ \bibnamefont
  {Mitchell}}\ and\ \bibinfo {author} {\bibfnamefont {W.~J.}\ \bibnamefont
  {Nellis}},\ }\href {https://doi.org/10.1063/1.329160} {\bibfield  {journal}
  {\bibinfo  {journal} {J. Appl. Phys.}\ }\textbf {\bibinfo {volume} {52}},\
  \bibinfo {pages} {3363} (\bibinfo {year} {1981})}\BibitemShut {NoStop}%
\bibitem [{\citenamefont {Knudson}\ \emph {et~al.}(2015)\citenamefont
  {Knudson}, \citenamefont {Desjarlais},\ and\ \citenamefont
  {Pribram-Jones}}]{knudson2015}%
  \BibitemOpen
  \bibfield  {author} {\bibinfo {author} {\bibfnamefont {M.~D.}\ \bibnamefont
  {Knudson}}, \bibinfo {author} {\bibfnamefont {M.~P.}\ \bibnamefont
  {Desjarlais}},\ and\ \bibinfo {author} {\bibfnamefont {A.}~\bibnamefont
  {Pribram-Jones}},\ }\href {https://doi.org/10.1103/PhysRevB.91.224105}
  {\bibfield  {journal} {\bibinfo  {journal} {Phys. Rev. B}\ }\textbf {\bibinfo
  {volume} {91}},\ \bibinfo {pages} {224105} (\bibinfo {year}
  {2015})}\BibitemShut {NoStop}%
\bibitem [{\citenamefont {Kerley}(1987)}]{kerley1987}%
  \BibitemOpen
  \bibfield  {author} {\bibinfo {author} {\bibfnamefont {G.~I.}\ \bibnamefont
  {Kerley}},\ }\href {https://doi.org/10.1016/0734-743X(87)90059-5} {\bibfield
  {journal} {\bibinfo  {journal} {Int. J. Impact Eng.}\ }\textbf {\bibinfo
  {volume} {5}},\ \bibinfo {pages} {441} (\bibinfo {year} {1987})}\BibitemShut
  {NoStop}%
\bibitem [{\citenamefont {Lomonosov}(2007)}]{lomonosov2007}%
  \BibitemOpen
  \bibfield  {author} {\bibinfo {author} {\bibfnamefont {I.~V.}\ \bibnamefont
  {Lomonosov}},\ }\href {https://doi.org/10.1017/S0263034607000687} {\bibfield
  {journal} {\bibinfo  {journal} {Laser Part. Beams}\ }\textbf {\bibinfo
  {volume} {25}},\ \bibinfo {pages} {567} (\bibinfo {year} {2007})}\BibitemShut
  {NoStop}%
\bibitem [{\citenamefont {Colgan}\ \emph {et~al.}(2015)\citenamefont {Colgan},
  \citenamefont {Kilcrease}, \citenamefont {Magee},\ and\ \citenamefont
  {{others}}}]{colgan2015}%
  \BibitemOpen
  \bibfield  {author} {\bibinfo {author} {\bibfnamefont {J.}~\bibnamefont
  {Colgan}}, \bibinfo {author} {\bibfnamefont {D.~P.}\ \bibnamefont
  {Kilcrease}}, \bibinfo {author} {\bibfnamefont {N.~H.}\ \bibnamefont
  {Magee}},\ and\ \bibinfo {author} {\bibnamefont {{others}}},\ }\href
  {https://doi.org/10.1016/j.hedp.2015.02.006} {\bibfield  {journal} {\bibinfo
  {journal} {High Energy Density Phys.}\ }\textbf {\bibinfo {volume} {14}},\
  \bibinfo {pages} {33} (\bibinfo {year} {2015})}\BibitemShut {NoStop}%
\bibitem [{\citenamefont {Recoules}\ and\ \citenamefont
  {Crocombette}(2005)}]{recoules2005}%
  \BibitemOpen
  \bibfield  {author} {\bibinfo {author} {\bibfnamefont {V.}~\bibnamefont
  {Recoules}}\ and\ \bibinfo {author} {\bibfnamefont {J.-P.}\ \bibnamefont
  {Crocombette}},\ }\href {https://doi.org/10.1103/PhysRevB.72.104202}
  {\bibfield  {journal} {\bibinfo  {journal} {Phys. Rev. B}\ }\textbf {\bibinfo
  {volume} {72}},\ \bibinfo {pages} {104202} (\bibinfo {year}
  {2005})}\BibitemShut {NoStop}%
\bibitem [{\citenamefont {Daligault}\ \emph {et~al.}(2016)\citenamefont
  {Daligault}, \citenamefont {Baalrud}, \citenamefont {Starrett}, \citenamefont
  {Saumon},\ and\ \citenamefont {Sjostrom}}]{daligault2016}%
  \BibitemOpen
  \bibfield  {author} {\bibinfo {author} {\bibfnamefont {J.}~\bibnamefont
  {Daligault}}, \bibinfo {author} {\bibfnamefont {S.~D.}\ \bibnamefont
  {Baalrud}}, \bibinfo {author} {\bibfnamefont {C.~E.}\ \bibnamefont
  {Starrett}}, \bibinfo {author} {\bibfnamefont {D.}~\bibnamefont {Saumon}},\
  and\ \bibinfo {author} {\bibfnamefont {T.}~\bibnamefont {Sjostrom}},\ }\href
  {https://doi.org/10.1103/PhysRevLett.116.075002} {\bibfield  {journal}
  {\bibinfo  {journal} {Phys. Rev. Lett.}\ }\textbf {\bibinfo {volume} {116}},\
  \bibinfo {pages} {075002} (\bibinfo {year} {2016})}\BibitemShut {NoStop}%
\bibitem [{\citenamefont {Sjostrom}\ \emph {et~al.}(2016)\citenamefont
  {Sjostrom}, \citenamefont {Crockett},\ and\ \citenamefont
  {Rudin}}]{sjostrom2016}%
  \BibitemOpen
  \bibfield  {author} {\bibinfo {author} {\bibfnamefont {T.}~\bibnamefont
  {Sjostrom}}, \bibinfo {author} {\bibfnamefont {S.}~\bibnamefont {Crockett}},\
  and\ \bibinfo {author} {\bibfnamefont {S.}~\bibnamefont {Rudin}},\ }\href
  {https://doi.org/10.1103/PhysRevB.94.144101} {\bibfield  {journal} {\bibinfo
  {journal} {Phys. Rev. B}\ }\textbf {\bibinfo {volume} {94}},\ \bibinfo
  {pages} {144101} (\bibinfo {year} {2016})}\BibitemShut {NoStop}%
\bibitem [{\citenamefont {Perry}\ \emph {et~al.}(2000)\citenamefont {Perry},
  \citenamefont {Davidson}, \citenamefont {Serduke}, \citenamefont {Bach},
  \citenamefont {Smith}, \citenamefont {Foster},\ and\ \citenamefont
  {{others}}}]{perry2000}%
  \BibitemOpen
  \bibfield  {author} {\bibinfo {author} {\bibfnamefont {T.~S.}\ \bibnamefont
  {Perry}}, \bibinfo {author} {\bibfnamefont {S.~J.}\ \bibnamefont {Davidson}},
  \bibinfo {author} {\bibfnamefont {F.~J.~D.}\ \bibnamefont {Serduke}},
  \bibinfo {author} {\bibfnamefont {D.~R.}\ \bibnamefont {Bach}}, \bibinfo
  {author} {\bibfnamefont {C.~C.}\ \bibnamefont {Smith}}, \bibinfo {author}
  {\bibfnamefont {J.~M.}\ \bibnamefont {Foster}},\ and\ \bibinfo {author}
  {\bibnamefont {{others}}},\ }\href {https://doi.org/10.1086/313368}
  {\bibfield  {journal} {\bibinfo  {journal} {Astrophys. J. Suppl. Ser.}\
  }\textbf {\bibinfo {volume} {127}},\ \bibinfo {pages} {433} (\bibinfo {year}
  {2000})}\BibitemShut {NoStop}%
\bibitem [{\citenamefont {Remington}\ \emph {et~al.}(2006)\citenamefont
  {Remington}, \citenamefont {Drake},\ and\ \citenamefont
  {Ryutov}}]{remington2006}%
  \BibitemOpen
  \bibfield  {author} {\bibinfo {author} {\bibfnamefont {B.~A.}\ \bibnamefont
  {Remington}}, \bibinfo {author} {\bibfnamefont {R.~P.}\ \bibnamefont
  {Drake}},\ and\ \bibinfo {author} {\bibfnamefont {D.~D.}\ \bibnamefont
  {Ryutov}},\ }\href {https://doi.org/10.1103/RevModPhys.78.755} {\bibfield
  {journal} {\bibinfo  {journal} {Rev. Mod. Phys.}\ }\textbf {\bibinfo {volume}
  {78}},\ \bibinfo {pages} {755} (\bibinfo {year} {2006})}\BibitemShut
  {NoStop}%
\bibitem [{\citenamefont {Gomez}\ and\ \citenamefont
  {{others}}(2014)}]{gomez2014}%
  \BibitemOpen
  \bibfield  {author} {\bibinfo {author} {\bibfnamefont {M.~R.}\ \bibnamefont
  {Gomez}}\ and\ \bibinfo {author} {\bibnamefont {{others}}},\ }\href
  {https://doi.org/10.1103/PhysRevLett.113.155003} {\bibfield  {journal}
  {\bibinfo  {journal} {Phys. Rev. Lett.}\ }\textbf {\bibinfo {volume} {113}},\
  \bibinfo {pages} {155003} (\bibinfo {year} {2014})}\BibitemShut {NoStop}%
\bibitem [{\citenamefont {Sinars}\ \emph {et~al.}(2010)\citenamefont {Sinars},
  \citenamefont {Slutz}, \citenamefont {Herrmann}, \citenamefont {McBride},
  \citenamefont {Cuneo}, \citenamefont {Peterson},\ and\ \citenamefont
  {{others}}}]{sinars2010}%
  \BibitemOpen
  \bibfield  {author} {\bibinfo {author} {\bibfnamefont {D.~B.}\ \bibnamefont
  {Sinars}}, \bibinfo {author} {\bibfnamefont {S.~A.}\ \bibnamefont {Slutz}},
  \bibinfo {author} {\bibfnamefont {M.~C.}\ \bibnamefont {Herrmann}}, \bibinfo
  {author} {\bibfnamefont {R.~D.}\ \bibnamefont {McBride}}, \bibinfo {author}
  {\bibfnamefont {M.~E.}\ \bibnamefont {Cuneo}}, \bibinfo {author}
  {\bibfnamefont {K.~J.}\ \bibnamefont {Peterson}},\ and\ \bibinfo {author}
  {\bibnamefont {{others}}},\ }\href
  {https://doi.org/10.1103/PhysRevLett.105.185001} {\bibfield  {journal}
  {\bibinfo  {journal} {Phys. Rev. Lett.}\ }\textbf {\bibinfo {volume} {105}},\
  \bibinfo {pages} {185001} (\bibinfo {year} {2010})}\BibitemShut {NoStop}%
\bibitem [{\citenamefont {Sinars}\ \emph {et~al.}(2011)\citenamefont {Sinars},
  \citenamefont {Slutz}, \citenamefont {Herrmann}, \citenamefont {McBride},
  \citenamefont {Cuneo}, \citenamefont {Jennings},\ and\ \citenamefont
  {{others}}}]{sinars2011}%
  \BibitemOpen
  \bibfield  {author} {\bibinfo {author} {\bibfnamefont {D.~B.}\ \bibnamefont
  {Sinars}}, \bibinfo {author} {\bibfnamefont {S.~A.}\ \bibnamefont {Slutz}},
  \bibinfo {author} {\bibfnamefont {M.~C.}\ \bibnamefont {Herrmann}}, \bibinfo
  {author} {\bibfnamefont {R.~D.}\ \bibnamefont {McBride}}, \bibinfo {author}
  {\bibfnamefont {M.~E.}\ \bibnamefont {Cuneo}}, \bibinfo {author}
  {\bibfnamefont {C.~A.}\ \bibnamefont {Jennings}},\ and\ \bibinfo {author}
  {\bibnamefont {{others}}},\ }\href {https://doi.org/10.1063/1.3560911}
  {\bibfield  {journal} {\bibinfo  {journal} {Phys. Plasmas}\ }\textbf {\bibinfo
  {volume} {18}},\ \bibinfo {pages} {056301} (\bibinfo {year}
  {2011})}\BibitemShut {NoStop}%
\bibitem [{\citenamefont {Kohn}\ and\ \citenamefont
  {Sham}(1965)}]{kohnsham1965}%
  \BibitemOpen
  \bibfield  {author} {\bibinfo {author} {\bibfnamefont {W.}~\bibnamefont
  {Kohn}}\ and\ \bibinfo {author} {\bibfnamefont {L.~J.}\ \bibnamefont
  {Sham}},\ }\href {https://doi.org/10.1103/PhysRev.140.A1133} {\bibfield
  {journal} {\bibinfo  {journal} {Phys. Rev.}\ }\textbf {\bibinfo {volume}
  {140}},\ \bibinfo {pages} {A1133} (\bibinfo {year} {1965})}\BibitemShut
  {NoStop}%
\bibitem [{\citenamefont {Mermin}(1965)}]{mermin1965}%
  \BibitemOpen
  \bibfield  {author} {\bibinfo {author} {\bibfnamefont {N.~D.}\ \bibnamefont
  {Mermin}},\ }\href {https://doi.org/10.1103/PhysRev.137.A1441} {\bibfield
  {journal} {\bibinfo  {journal} {Phys. Rev.}\ }\textbf {\bibinfo {volume}
  {137}},\ \bibinfo {pages} {A1441} (\bibinfo {year} {1965})}\BibitemShut
  {NoStop}%
\bibitem [{\citenamefont {Desjarlais}\ \emph {et~al.}(2002)\citenamefont
  {Desjarlais}, \citenamefont {Kress},\ and\ \citenamefont
  {Collins}}]{desjarlais2002}%
  \BibitemOpen
  \bibfield  {author} {\bibinfo {author} {\bibfnamefont {M.~P.}\ \bibnamefont
  {Desjarlais}}, \bibinfo {author} {\bibfnamefont {J.~D.}\ \bibnamefont
  {Kress}},\ and\ \bibinfo {author} {\bibfnamefont {L.~A.}\ \bibnamefont
  {Collins}},\ }\href {https://doi.org/10.1103/PhysRevE.66.025401} {\bibfield
  {journal} {\bibinfo  {journal} {Phys. Rev. E}\ }\textbf {\bibinfo {volume}
  {66}},\ \bibinfo {pages} {025401} (\bibinfo {year} {2002})}\BibitemShut
  {NoStop}%
\bibitem [{\citenamefont {Vo{\v{c}}adlo}\ and\ \citenamefont
  {Alf{\`e}}(2002)}]{vocadlo2002}%
  \BibitemOpen
  \bibfield  {author} {\bibinfo {author} {\bibfnamefont {L.}~\bibnamefont
  {Vo{\v{c}}adlo}}\ and\ \bibinfo {author} {\bibfnamefont {D.}~\bibnamefont
  {Alf{\`e}}},\ }\href {https://doi.org/10.1103/PhysRevB.65.214105} {\bibfield
  {journal} {\bibinfo  {journal} {Phys. Rev. B}\ }\textbf {\bibinfo {volume}
  {65}},\ \bibinfo {pages} {214105} (\bibinfo {year} {2002})}\BibitemShut
  {NoStop}%
\bibitem [{\citenamefont {Mazevet}\ \emph {et~al.}(2005)\citenamefont
  {Mazevet}, \citenamefont {Desjarlais}, \citenamefont {Collins}, \citenamefont
  {Kress},\ and\ \citenamefont {Magee}}]{mazewet2005}%
  \BibitemOpen
  \bibfield  {author} {\bibinfo {author} {\bibfnamefont {S.}~\bibnamefont
  {Mazevet}}, \bibinfo {author} {\bibfnamefont {M.~P.}\ \bibnamefont
  {Desjarlais}}, \bibinfo {author} {\bibfnamefont {L.~A.}\ \bibnamefont
  {Collins}}, \bibinfo {author} {\bibfnamefont {J.~D.}\ \bibnamefont {Kress}},\
  and\ \bibinfo {author} {\bibfnamefont {N.~H.}\ \bibnamefont {Magee}},\ }\href
  {https://doi.org/10.1103/PhysRevE.71.016409} {\bibfield  {journal} {\bibinfo
  {journal} {Phys. Rev. E}\ }\textbf {\bibinfo {volume} {71}},\ \bibinfo
  {pages} {016409} (\bibinfo {year} {2005})}\BibitemShut {NoStop}%
\bibitem [{\citenamefont {Bouchet}\ \emph {et~al.}(2009)\citenamefont
  {Bouchet}, \citenamefont {Bottin}, \citenamefont {Jomard},\ and\
  \citenamefont {Z{\'e}rah}}]{bouchet2009}%
  \BibitemOpen
  \bibfield  {author} {\bibinfo {author} {\bibfnamefont {J.}~\bibnamefont
  {Bouchet}}, \bibinfo {author} {\bibfnamefont {F.}~\bibnamefont {Bottin}},
  \bibinfo {author} {\bibfnamefont {G.}~\bibnamefont {Jomard}},\ and\ \bibinfo
  {author} {\bibfnamefont {G.}~\bibnamefont {Z{\'e}rah}},\ }\href
  {https://doi.org/10.1103/PhysRevB.80.094102} {\bibfield  {journal} {\bibinfo
  {journal} {Phys. Rev. B}\ }\textbf {\bibinfo {volume} {80}},\ \bibinfo
  {pages} {094102} (\bibinfo {year} {2009})}\BibitemShut {NoStop}%
\bibitem [{\citenamefont {Vl{\v{c}}ek}\ \emph {et~al.}(2012)\citenamefont
  {Vl{\v{c}}ek}, \citenamefont {de~Koker},\ and\ \citenamefont
  {Steinle-Neumann}}]{vlcek2012}%
  \BibitemOpen
  \bibfield  {author} {\bibinfo {author} {\bibfnamefont {V.}~\bibnamefont
  {Vl{\v{c}}ek}}, \bibinfo {author} {\bibfnamefont {N.}~\bibnamefont
  {de~Koker}},\ and\ \bibinfo {author} {\bibfnamefont {G.}~\bibnamefont
  {Steinle-Neumann}},\ }\href {https://doi.org/10.1103/PhysRevB.85.184201}
  {\bibfield  {journal} {\bibinfo  {journal} {Phys. Rev. B}\ }\textbf {\bibinfo
  {volume} {85}},\ \bibinfo {pages} {184201} (\bibinfo {year}
  {2012})}\BibitemShut {NoStop}%
\bibitem [{\citenamefont {Minakov}\ \emph {et~al.}(2014)\citenamefont
  {Minakov}, \citenamefont {Levashov}, \citenamefont {Khishchenko},\ and\
  \citenamefont {Fortov}}]{minakov2014}%
  \BibitemOpen
  \bibfield  {author} {\bibinfo {author} {\bibfnamefont {D.~V.}\ \bibnamefont
  {Minakov}}, \bibinfo {author} {\bibfnamefont {P.~R.}\ \bibnamefont
  {Levashov}}, \bibinfo {author} {\bibfnamefont {K.~V.}\ \bibnamefont
  {Khishchenko}},\ and\ \bibinfo {author} {\bibfnamefont {V.~E.}\ \bibnamefont
  {Fortov}},\ }\href {https://doi.org/10.1063/1.4882299} {\bibfield  {journal}
  {\bibinfo  {journal} {J. Appl. Phys.}\ }\textbf {\bibinfo {volume} {115}},\
  \bibinfo {pages} {223512} (\bibinfo {year} {2014})}\BibitemShut {NoStop}%
\bibitem [{\citenamefont {Witte}\ \emph {et~al.}(2017)\citenamefont {Witte},
  \citenamefont {Fletcher}, \citenamefont {Galtier}, \citenamefont {Gamboa},
  \citenamefont {Lee}, \citenamefont {Zastrau},\ and\ \citenamefont
  {{others}}}]{witte2017}%
  \BibitemOpen
  \bibfield  {author} {\bibinfo {author} {\bibfnamefont {B.~B.~L.}\
  \bibnamefont {Witte}}, \bibinfo {author} {\bibfnamefont {L.~B.}\ \bibnamefont
  {Fletcher}}, \bibinfo {author} {\bibfnamefont {E.}~\bibnamefont {Galtier}},
  \bibinfo {author} {\bibfnamefont {E.~J.}\ \bibnamefont {Gamboa}}, \bibinfo
  {author} {\bibfnamefont {H.~J.}\ \bibnamefont {Lee}}, \bibinfo {author}
  {\bibfnamefont {U.}~\bibnamefont {Zastrau}},\ and\ \bibinfo {author}
  {\bibnamefont {{others}}},\ }\href
  {https://doi.org/10.1103/PhysRevLett.118.225001} {\bibfield  {journal}
  {\bibinfo  {journal} {Phys. Rev. Lett.}\ }\textbf {\bibinfo {volume} {118}},\
  \bibinfo {pages} {225001} (\bibinfo {year} {2017})}\BibitemShut {NoStop}%
\bibitem [{\citenamefont {Blanchet}\ \emph {et~al.}(2020)\citenamefont
  {Blanchet}, \citenamefont {Torrent},\ and\ \citenamefont
  {Cl{\'e}rouin}}]{blanchet2020}%
  \BibitemOpen
  \bibfield  {author} {\bibinfo {author} {\bibfnamefont {A.}~\bibnamefont
  {Blanchet}}, \bibinfo {author} {\bibfnamefont {M.}~\bibnamefont {Torrent}},\
  and\ \bibinfo {author} {\bibfnamefont {J.}~\bibnamefont {Cl{\'e}rouin}},\
  }\href {https://doi.org/10.1063/5.0016538} {\bibfield  {journal} {\bibinfo
  {journal} {Phys. Plasmas}\ }\textbf {\bibinfo {volume} {27}},\ \bibinfo
  {pages} {122706} (\bibinfo {year} {2020})}\BibitemShut {NoStop}%
\bibitem [{\citenamefont {More}\ \emph {et~al.}(1988)\citenamefont {More},
  \citenamefont {Warren}, \citenamefont {Young},\ and\ \citenamefont
  {Zimmerman}}]{more1988}%
  \BibitemOpen
  \bibfield  {author} {\bibinfo {author} {\bibfnamefont {R.~M.}\ \bibnamefont
  {More}}, \bibinfo {author} {\bibfnamefont {K.~H.}\ \bibnamefont {Warren}},
  \bibinfo {author} {\bibfnamefont {D.~A.}\ \bibnamefont {Young}},\ and\
  \bibinfo {author} {\bibfnamefont {G.~B.}\ \bibnamefont {Zimmerman}},\ }\href
  {https://doi.org/10.1063/1.866963} {\bibfield  {journal} {\bibinfo  {journal}
  {Phys. Fluids}\ }\textbf {\bibinfo {volume} {31}},\ \bibinfo {pages} {3059}
  (\bibinfo {year} {1988})}\BibitemShut {NoStop}%
\bibitem [{\citenamefont {MacFarlane}\ \emph {et~al.}(2006)\citenamefont
  {MacFarlane}, \citenamefont {Golovkin},\ and\ \citenamefont
  {Woodruff}}]{macfarlane2006}%
  \BibitemOpen
  \bibfield  {author} {\bibinfo {author} {\bibfnamefont {J.~J.}\ \bibnamefont
  {MacFarlane}}, \bibinfo {author} {\bibfnamefont {I.~E.}\ \bibnamefont
  {Golovkin}},\ and\ \bibinfo {author} {\bibfnamefont {P.~R.}\ \bibnamefont
  {Woodruff}},\ }\href {https://doi.org/10.1016/j.jqsrt.2005.05.031} {\bibfield
   {journal} {\bibinfo  {journal} {J. Quant. Spectrosc. Radiat. Transfer}\
  }\textbf {\bibinfo {volume} {99}},\ \bibinfo {pages} {381} (\bibinfo {year}
  {2006})}\BibitemShut {NoStop}%
\bibitem [{\citenamefont {Spitzer}\ and\ \citenamefont
  {H{\"a}rm}(1953)}]{spitzer1953}%
  \BibitemOpen
  \bibfield  {author} {\bibinfo {author} {\bibfnamefont {L.}~\bibnamefont
  {Spitzer}}\ and\ \bibinfo {author} {\bibfnamefont {R.}~\bibnamefont
  {H{\"a}rm}},\ }\href {https://doi.org/10.1103/PhysRev.89.977} {\bibfield
  {journal} {\bibinfo  {journal} {Phys. Rev.}\ }\textbf {\bibinfo {volume}
  {89}},\ \bibinfo {pages} {977} (\bibinfo {year} {1953})}\BibitemShut
  {NoStop}%
\bibitem [{\citenamefont {Lee}\ and\ \citenamefont {More}(1984)}]{leemore1984}%
  \BibitemOpen
  \bibfield  {author} {\bibinfo {author} {\bibfnamefont {Y.~T.}\ \bibnamefont
  {Lee}}\ and\ \bibinfo {author} {\bibfnamefont {R.~M.}\ \bibnamefont {More}},\
  }\href {https://doi.org/10.1063/1.864744} {\bibfield  {journal} {\bibinfo
  {journal} {Phys. Fluids}\ }\textbf {\bibinfo {volume} {27}},\ \bibinfo
  {pages} {1273} (\bibinfo {year} {1984})}\BibitemShut {NoStop}%
\bibitem [{\citenamefont {Cytter}\ \emph {et~al.}(2018)\citenamefont {Cytter},
  \citenamefont {Rabani}, \citenamefont {Neuhauser},\ and\ \citenamefont
  {Baer}}]{cytter2018}%
  \BibitemOpen
  \bibfield  {author} {\bibinfo {author} {\bibfnamefont {Y.}~\bibnamefont
  {Cytter}}, \bibinfo {author} {\bibfnamefont {E.}~\bibnamefont {Rabani}},
  \bibinfo {author} {\bibfnamefont {D.}~\bibnamefont {Neuhauser}},\ and\
  \bibinfo {author} {\bibfnamefont {R.}~\bibnamefont {Baer}},\ }\href
  {https://doi.org/10.1103/PhysRevB.97.115207} {\bibfield  {journal} {\bibinfo
  {journal} {Phys. Rev. B}\ }\textbf {\bibinfo {volume} {97}},\ \bibinfo
  {pages} {115207} (\bibinfo {year} {2018})}\BibitemShut {NoStop}%
\bibitem [{\citenamefont {White}\ and\ \citenamefont
  {Collins}(2020)}]{white2020}%
  \BibitemOpen
  \bibfield  {author} {\bibinfo {author} {\bibfnamefont {A.~J.}\ \bibnamefont
  {White}}\ and\ \bibinfo {author} {\bibfnamefont {L.~A.}\ \bibnamefont
  {Collins}},\ }\href {https://doi.org/10.1103/PhysRevLett.125.055002}
  {\bibfield  {journal} {\bibinfo  {journal} {Phys. Rev. Lett.}\ }\textbf
  {\bibinfo {volume} {125}},\ \bibinfo {pages} {055002} (\bibinfo {year}
  {2020})}\BibitemShut {NoStop}%
\bibitem [{\citenamefont {Liu}\ \emph {et~al.}(2025)\citenamefont {Liu},
  \citenamefont {He},\ and\ \citenamefont {Chen}}]{liu2025altransport}%
  \BibitemOpen
  \bibfield  {author} {\bibinfo {author} {\bibfnamefont {Q.}~\bibnamefont
  {Liu}}, \bibinfo {author} {\bibfnamefont {X.}~\bibnamefont {He}},\ and\
  \bibinfo {author} {\bibfnamefont {M.}~\bibnamefont {Chen}},\ }\href@noop {}
  {\bibfield  {journal} {\bibinfo  {journal} {arXiv preprint arXiv:2510.10112}\
  } (\bibinfo {year} {2025})},\ \Eprint {https://arxiv.org/abs/2510.10112}
  {arXiv:2510.10112 [cond-mat.mtrl-sci]} \BibitemShut {NoStop}%
\bibitem [{\citenamefont {Liu}\ and\ \citenamefont {Chen}(2022)}]{liu2022}%
  \BibitemOpen
  \bibfield  {author} {\bibinfo {author} {\bibfnamefont {Q.}~\bibnamefont
  {Liu}}\ and\ \bibinfo {author} {\bibfnamefont {M.}~\bibnamefont {Chen}},\
  }\href {https://doi.org/10.1103/PhysRevB.106.125132} {\bibfield  {journal}
  {\bibinfo  {journal} {Phys. Rev. B}\ }\textbf {\bibinfo {volume} {106}},\
  \bibinfo {pages} {125132} (\bibinfo {year} {2022})}\BibitemShut {NoStop}%
\bibitem [{\citenamefont {Hamann}(2013)}]{hamann2013}%
  \BibitemOpen
  \bibfield  {author} {\bibinfo {author} {\bibfnamefont {D.~R.}\ \bibnamefont
  {Hamann}},\ }\href {https://doi.org/10.1103/PhysRevB.88.085117} {\bibfield
  {journal} {\bibinfo  {journal} {Phys. Rev. B}\ }\textbf {\bibinfo {volume}
  {88}},\ \bibinfo {pages} {085117} (\bibinfo {year} {2013})}\BibitemShut
  {NoStop}%
\bibitem [{\citenamefont {Perdew}\ \emph {et~al.}(1996)\citenamefont {Perdew},
  \citenamefont {Burke},\ and\ \citenamefont {Ernzerhof}}]{perdew1996}%
  \BibitemOpen
  \bibfield  {author} {\bibinfo {author} {\bibfnamefont {J.~P.}\ \bibnamefont
  {Perdew}}, \bibinfo {author} {\bibfnamefont {K.}~\bibnamefont {Burke}},\ and\
  \bibinfo {author} {\bibfnamefont {M.}~\bibnamefont {Ernzerhof}},\ }\href
  {https://doi.org/10.1103/PhysRevLett.77.3865} {\bibfield  {journal} {\bibinfo
   {journal} {Phys. Rev. Lett.}\ }\textbf {\bibinfo {volume} {77}},\ \bibinfo
  {pages} {3865} (\bibinfo {year} {1996})}\BibitemShut {NoStop}%
\bibitem [{\citenamefont {Duvall}\ and\ \citenamefont
  {Graham}(1977)}]{duvall1977}%
  \BibitemOpen
  \bibfield  {author} {\bibinfo {author} {\bibfnamefont {G.~E.}\ \bibnamefont
  {Duvall}}\ and\ \bibinfo {author} {\bibfnamefont {R.~A.}\ \bibnamefont
  {Graham}},\ }\href {https://doi.org/10.1103/RevModPhys.49.523} {\bibfield
  {journal} {\bibinfo  {journal} {Rev. Mod. Phys.}\ }\textbf {\bibinfo
  {volume} {49}},\ \bibinfo {pages} {523} (\bibinfo {year}
  {1977})}\BibitemShut {NoStop}%
\bibitem [{\citenamefont {Simonenko}\ \emph {et~al.}(1985)\citenamefont
  {Simonenko}, \citenamefont {Voloshin}, \citenamefont {Vladimirov},
  \citenamefont {Nagibin}, \citenamefont {Nogin}, \citenamefont {Popov},
  \citenamefont {Vasilenko},\ and\ \citenamefont {Shoidin}}]{simonenko1985}%
  \BibitemOpen
  \bibfield  {author} {\bibinfo {author} {\bibfnamefont {V.~A.}\ \bibnamefont
  {Simonenko}}, \bibinfo {author} {\bibfnamefont {N.~P.}\ \bibnamefont
  {Voloshin}}, \bibinfo {author} {\bibfnamefont {A.~S.}\ \bibnamefont
  {Vladimirov}}, \bibinfo {author} {\bibfnamefont {A.~P.}\ \bibnamefont
  {Nagibin}}, \bibinfo {author} {\bibfnamefont {V.~N.}\ \bibnamefont {Nogin}},
  \bibinfo {author} {\bibfnamefont {V.~A.}\ \bibnamefont {Popov}}, \bibinfo
  {author} {\bibfnamefont {V.~A.}\ \bibnamefont {Vasilenko}},\ and\ \bibinfo
  {author} {\bibfnamefont {Y.~A.}\ \bibnamefont {Shoidin}},\ }\href@noop {}
  {\bibfield  {journal} {\bibinfo  {journal} {Sov. Phys. JETP}\ }\textbf
  {\bibinfo {volume} {61}},\ \bibinfo {pages} {869} (\bibinfo {year}
  {1985})}\BibitemShut {NoStop}%
\bibitem [{\citenamefont {Ragan}(1982)}]{ragan1982}%
  \BibitemOpen
  \bibfield  {author} {\bibinfo {author} {\bibfnamefont {C.~E.}\ \bibnamefont
  {Ragan}, \bibfnamefont {III}},\ }\href
  {https://doi.org/10.1103/PhysRevA.25.3360} {\bibfield  {journal} {\bibinfo
  {journal} {Phys. Rev. A}\ }\textbf {\bibinfo {volume} {25}},\ \bibinfo
  {pages} {3360} (\bibinfo {year} {1982})}\BibitemShut {NoStop}%
\bibitem [{\citenamefont {Avrorin}\ \emph {et~al.}(1987)\citenamefont
  {Avrorin}, \citenamefont {Vodolaga}, \citenamefont {Voloshin}, \citenamefont
  {Kovalenko}, \citenamefont {Kuropatenko}, \citenamefont {Simonenko},\ and\
  \citenamefont {Chernovolyuk}}]{avrorin1987}%
  \BibitemOpen
  \bibfield  {author} {\bibinfo {author} {\bibfnamefont {E.~N.}\ \bibnamefont
  {Avrorin}}, \bibinfo {author} {\bibfnamefont {B.~K.}\ \bibnamefont
  {Vodolaga}}, \bibinfo {author} {\bibfnamefont {N.~P.}\ \bibnamefont
  {Voloshin}}, \bibinfo {author} {\bibfnamefont {G.~V.}\ \bibnamefont
  {Kovalenko}}, \bibinfo {author} {\bibfnamefont {V.~F.}\ \bibnamefont
  {Kuropatenko}}, \bibinfo {author} {\bibfnamefont {V.~A.}\ \bibnamefont
  {Simonenko}},\ and\ \bibinfo {author} {\bibfnamefont {B.~T.}\ \bibnamefont
  {Chernovolyuk}},\ }\href@noop {} {\bibfield  {journal} {\bibinfo  {journal}
  {Sov. Phys. JETP}\ }\textbf {\bibinfo {volume} {66}},\ \bibinfo {pages} {347}
  (\bibinfo {year} {1987})}\BibitemShut {NoStop}%
\bibitem [{\citenamefont {Ragan}(1984)}]{ragan1984}%
  \BibitemOpen
  \bibfield  {author} {\bibinfo {author} {\bibfnamefont {C.~E.}\ \bibnamefont
  {Ragan}, \bibfnamefont {III}},\ }\href
  {https://doi.org/10.1103/PhysRevA.29.1391} {\bibfield  {journal} {\bibinfo
  {journal} {Phys. Rev. A}\ }\textbf {\bibinfo {volume} {29}},\ \bibinfo
  {pages} {1391} (\bibinfo {year} {1984})}\BibitemShut {NoStop}%
\bibitem [{\citenamefont {Trunin}\ \emph
  {et~al.}(1995{\natexlab{a}})\citenamefont {Trunin}, \citenamefont {Podurets},
  \citenamefont {Simakov}, \citenamefont {Popov},\ and\ \citenamefont
  {Sevast'yanov}}]{trunin1995jetp}%
  \BibitemOpen
  \bibfield  {author} {\bibinfo {author} {\bibfnamefont {R.~F.}\ \bibnamefont
  {Trunin}}, \bibinfo {author} {\bibfnamefont {M.~A.}\ \bibnamefont
  {Podurets}}, \bibinfo {author} {\bibfnamefont {G.~V.}\ \bibnamefont
  {Simakov}}, \bibinfo {author} {\bibfnamefont {L.~V.}\ \bibnamefont {Popov}},\
  and\ \bibinfo {author} {\bibfnamefont {A.~G.}\ \bibnamefont {Sevast'yanov}},\
  }\href@noop {} {\bibfield  {journal} {\bibinfo  {journal} {J. Exp. Theor.
  Phys.}\ }\textbf {\bibinfo {volume} {81}},\ \bibinfo {pages} {464} (\bibinfo
  {year} {1995}{\natexlab{a}})}\BibitemShut {NoStop}%
\bibitem [{\citenamefont {Trunin}\ \emph
  {et~al.}(1995{\natexlab{b}})\citenamefont {Trunin}, \citenamefont {Panov},\
  and\ \citenamefont {Medvedev}}]{trunin1995ht}%
  \BibitemOpen
  \bibfield  {author} {\bibinfo {author} {\bibfnamefont {R.~F.}\ \bibnamefont
  {Trunin}}, \bibinfo {author} {\bibfnamefont {N.~V.}\ \bibnamefont {Panov}},\
  and\ \bibinfo {author} {\bibfnamefont {A.~B.}\ \bibnamefont {Medvedev}},\
  }\href@noop {} {\bibfield  {journal} {\bibinfo  {journal} {High Temp.}\
  }\textbf {\bibinfo {volume} {33}},\ \bibinfo {pages} {328} (\bibinfo {year}
  {1995}{\natexlab{b}})}\BibitemShut {NoStop}%
\bibitem [{\citenamefont {Epperlein}\ and\ \citenamefont
  {Haines}(1986)}]{epperlein1986}%
  \BibitemOpen
  \bibfield  {author} {\bibinfo {author} {\bibfnamefont {E.~M.}\ \bibnamefont
  {Epperlein}}\ and\ \bibinfo {author} {\bibfnamefont {M.~G.}\ \bibnamefont
  {Haines}},\ }\href {https://doi.org/10.1063/1.865670} {\bibfield  {journal}
  {\bibinfo  {journal} {Phys. Fluids}\ }\textbf {\bibinfo {volume} {29}},\
  \bibinfo {pages} {1029} (\bibinfo {year} {1986})}\BibitemShut {NoStop}%
\bibitem [{\citenamefont {Braginskii}(1958)}]{braginskii1958}%
  \BibitemOpen
  \bibfield  {author} {\bibinfo {author} {\bibfnamefont {S.~I.}\ \bibnamefont
  {Braginskii}},\ }\href@noop {} {\bibfield  {journal} {\bibinfo
  {journal} {Sov. Phys. JETP}\ }\textbf {\bibinfo {volume}
  {6}},\ \bibinfo {pages} {358} (\bibinfo {year} {1958})}\BibitemShut
  {NoStop}%
\bibitem [{\citenamefont {Wu}\ \emph {et~al.}(2021)\citenamefont {Wu},
  \citenamefont {Myint}, \citenamefont {Pask}, \citenamefont {Prisbrey},
  \citenamefont {Correa}, \citenamefont {Suryanarayana},\ and\ \citenamefont
  {Varley}}]{wu2021}%
  \BibitemOpen
  \bibfield  {author} {\bibinfo {author} {\bibfnamefont {C.~J.}\ \bibnamefont
  {Wu}}, \bibinfo {author} {\bibfnamefont {P.~C.}\ \bibnamefont {Myint}},
  \bibinfo {author} {\bibfnamefont {J.~E.}\ \bibnamefont {Pask}}, \bibinfo
  {author} {\bibfnamefont {C.~J.}\ \bibnamefont {Prisbrey}}, \bibinfo {author}
  {\bibfnamefont {A.~A.}\ \bibnamefont {Correa}}, \bibinfo {author}
  {\bibfnamefont {P.}~\bibnamefont {Suryanarayana}},\ and\ \bibinfo {author}
  {\bibfnamefont {J.~B.}\ \bibnamefont {Varley}},\ }\href
  {https://doi.org/10.1021/acs.jpca.0c09809} {\bibfield  {journal} {\bibinfo
  {journal} {J. Phys. Chem. A}\ }\textbf {\bibinfo {volume} {125}},\ \bibinfo
  {pages} {1610} (\bibinfo {year} {2021})}\BibitemShut {NoStop}%
\bibitem [{\citenamefont {Vinet}\ \emph {et~al.}(1989)\citenamefont {Vinet},
  \citenamefont {Rose}, \citenamefont {Ferrante},\ and\ \citenamefont
  {Smith}}]{vinet1989}%
  \BibitemOpen
  \bibfield  {author} {\bibinfo {author} {\bibfnamefont {P.}~\bibnamefont
  {Vinet}}, \bibinfo {author} {\bibfnamefont {J.~H.}\ \bibnamefont {Rose}},
  \bibinfo {author} {\bibfnamefont {J.}~\bibnamefont {Ferrante}},\ and\
  \bibinfo {author} {\bibfnamefont {J.~R.}\ \bibnamefont {Smith}},\ }\href
  {https://doi.org/10.1088/0953-8984/1/11/002} {\bibfield  {journal} {\bibinfo
  {journal} {J. Phys.: Condens. Matter}\ }\textbf {\bibinfo {volume} {1}},\
  \bibinfo {pages} {1941} (\bibinfo {year} {1989})}\BibitemShut {NoStop}%
\end{thebibliography}
\end{document}